\documentclass[11pt]{article}

\usepackage{amsfonts,amsthm}
\usepackage{amsmath,mathtools}   
\usepackage{latexsym}
\usepackage{amssymb}
\usepackage{mathrsfs}
\usepackage{pifont}
\usepackage{tikz,ulem}
\usepackage[all]{xy}
\usepackage{hyperref}

\newtheoremstyle{newrem}{3pt}{3pt}{}{}
{\bfseries}{.}{.5em}{}

\newtheorem{theo}{Theorem}[section]

\newtheorem*{theo*}{Theorem}

\newtheorem{prop}[theo]{Proposition}

\theoremstyle{newrem}

\theoremstyle{definition}

\newtheorem*{term*}{Notation/Terminology}

\newcommand{\cW}{\mathcal{W}}

\def\sk{\mathsf{k}}    \def\sw{\mathsf{w}}    \def\sh{\mathsf{h}}    
    \def\sG{\mathsf{G}}   \def\sT{\mathsf{T}}    
\def\sM{\mathsf{M}}

\newcommand{\II}{{\mathbb I}}
\newcommand{\ZZ}{{\mathbb Z}}

\newcommand{\bleu}[1]{\textcolor{blue}{#1}}

\newcommand{\bk}{\boldsymbol{\sk}}
\newcommand{\bh}{\boldsymbol{\sh}}
\newcommand{\bw}{\boldsymbol{\sw}}
\newcommand{\bT}{\boldsymbol{\sT}}
\newcommand{\bG}{\boldsymbol{\sG}}
\newcommand{\bM}{\boldsymbol{\sM}}

\usepackage{geometry}
\newcommand{\doi}[1]{\href{https://doi.org/#1}{doi:#1}}
\numberwithin{equation}{section}

\begin{document}

\title{\bf $\lambda$-Griffiths polynomials:\\ Bispectrality and biorthogonality\\
}
\author{
N. Cramp\'e\textsuperscript{$1$}\footnote{E-mail: crampe1977@gmail.com}~,
L. Frappat\textsuperscript{$2$}\footnote{E-mail: luc.frappat@lapth.cnrs.fr}~,
J. Gaboriaud\textsuperscript{$3$}\footnote{E-mail: julien.gaboriaud.57f@st.kyoto-u.ac.jp}~,
E. Ragoucy\textsuperscript{$2$}\footnote{E-mail: eric.ragoucy@lapth.cnrs.fr}~,
L. Vinet\textsuperscript{$4,5$}\footnote{E-mail: luc.vinet@umontreal.ca}~,
M. Zaimi\textsuperscript{$4$}\footnote{E-mail: meri.zaimi@umontreal.ca}~,
\\[.9em]
\textsuperscript{$1$}
\small Institut Denis-Poisson CNRS/UMR 7013 - Universit\'e de Tours - Universit\'e
d'Orl\'eans,\\
\small~Parc de Grandmont, 37200 Tours, France.\\[.5em]
\textsuperscript{$2$}
\small Laboratoire d'Annecy-le-Vieux de Physique Th\'eorique LAPTh,\\
\small~Universit\'e Savoie Mont Blanc, CNRS, F-74000 Annecy,
 France.\\[.9em]
 \textsuperscript{$3$}
\small Graduate School of Informatics, Kyoto University, Sakyo-ku, Kyoto, 606-8501, Japan.\\[.9em]
 \textsuperscript{$4$}
\small Centre de Recherches Math\'ematiques, Universit\'e de Montr\'eal, P.O. Box 6128, \\
\small Centre-ville Station, Montr\'eal (Qu\'ebec), H3C 3J7, Canada.\\[.9em]
\textsuperscript{$5$}
\small IVADO, Montr\'eal (Qu\'ebec), H2S 3H1, Canada.\\[.9em]
}
\date{}
\maketitle

\bigskip\bigskip 

\begin{center}
\begin{minipage}{14cm}
\begin{center}
{\bf Abstract}\\
\end{center}
We introduce a generalization of bivariate Griffiths polynomials depending on an additional parameter $\lambda$. 
These $\lambda$-Griffiths polynomials are bivariate, bispectral and biorthogonal. For two specific values 
of the parameter $\lambda$, they become orthogonal. One of the value is related to the usual bivariate Griffiths polynomials, 
while the second value produces new orthogonal bivariate polynomials.
\end{minipage}
\end{center}

\medskip

\begin{center}
\begin{minipage}{14cm}

\textbf{MSC2020 database:} 33C80; 33C45; 16G60
\end{minipage}
\end{center}

\clearpage
\newpage

\section{Introduction}\label{sec:intro}

The bispectral problem involves two linear operators $L_n$ and $L_x$
which act on the variables $n$ and $x$ respectively and have the same eigenfunctions. 
More precisely, the differential or difference operator $L_n$ is independent of the variable $x$ and acts on functions of $n$, and similarly 
the differential or difference operator 
$L_x$
is independent of the variable $n$ and acts on functions of $x$. The functions $\psi(n,x)$ are eigenfunctions of both operators:
\begin{align}
 L_n \psi(n,x)= \mu(x)\psi(n,x)\,,\nonumber\\
  L_x \psi(n,x)= \nu(n)\psi(n,x)\,,\nonumber
\end{align}
where $\mu(x)$ is a function of $x$ only and $\nu(n)$ of $n$ only. 
This problem leads to exciting subjects of exploration across multiple disciplines, including 
integrable systems, quantum mechanics, and differential equations. 
The case where both operators are differential operators has been investigated in \cite{DG}. The cases where $L_n$ becomes a three-term difference operator and $L_x$ is a second order differential operator 
or a three-term difference operator is related to the study of the orthogonal polynomials of the Askey scheme. This paper aims to generalize 
 these results for bivariate polynomials.

The generalizations with more than one variable of the Krawtchouk polynomials, introduced in \cite{Griff} (see also \cite{DC}) and called Griffiths polynomials, are orthogonal polynomials with respect to the multinomial distribution.
These polynomials in the bivariate case were rediscovered in 2008 in relation to a probabilistic model and as limits of the $su(2)$  $9j$-symbols \cite{HR} (they were called Rahman polynomials for a while). 
In \cite{MT}, they have been related to character algebras and an explicit formula in terms of Gel’fand-Aomoto hypergeometric series was given.
They also showed up in the context of the Racah problem for the oscillator algebra \cite{Zhe,CVV}.
Moreover, these polynomials appear as overlap coefficients between two modules of $sl_{n+1}(\mathbb{C})$ with their basis elements defined as eigenvectors of two Cartan 
subalgebras related by an antiautomorphism depending on the parameters entering the polynomials \cite{Ros,IT,I12}.
Closely related to the last interpretation, they arise in the matrix elements of the orthogonal group $O(n+ 1)$ representations
that act on the energy eigenspaces of the isotropic $(n+ 1)$-dimensional harmonic oscillator \cite{GVZ}. 
This last interpretation provides a proof of the bispectrality property for the Griffiths polynomials.

By fixing some parameters in the Griffiths polynomials, they become the so-called Tratnik polynomials of Krawtchouk type \cite{Trat}. 
These latter exist for the other polynomials of the Askey scheme and remain bispectral \cite{GI,I12}. At this moment, the equivalent scheme is still missing 
for Griffiths-like polynomials. Let us mention some recent advances in this direction for a generalization of the bivariate Griffiths polynomials, which is 
called bivariate Griffiths polynomials of Racah type \cite{icosi}. 
To try to fill this gap, we obtain in this paper the bispectrality of the bivariate Griffiths polynomials by using only properties of the univariate polynomials.
We hope that this elementary method can be expanded to other polynomials.

Thanks to this new approach, we get a generalization of the bivariate Griffiths polynomials depending on an additional parameter $\lambda$, that we call
 $\lambda$-Griffiths polynomials.  They are still bispectral.
However, in opposition to the usual Griffiths polynomials, these polynomials are not orthogonal anymore, but biorthogonal. The notion of biorthogonality has already appeared in the study 
of multivariate polynomials, see \textit{e.g.}\ \cite{Trat89,Trat}. It also arises more generally in connection to rational functions \cite{Wil91,IM,GM} and provides another natural direction 
for extending the Askey scheme of orthogonal polynomials. From an algebraic perspective, biorthogonal functions occur as solutions of generalized eigenvalue problems of the form 
$L \psi = \mu M \psi$; when the operators $L$ and $M$ act tridiagonally in a certain basis, the functions $\psi$ are rational (or polynomial) \cite{Zhe99}. 
Then, bispectrality in this context is obtained when considering two such generalized eigenvalue problems (see the case of the Hahn biorthogonal rational functions treated in \cite{TVZ,VZ_Hahn} and various other cases in \cite{VZ_Ask,BGVZ,VZZ,CVZZ}).

The plan of this paper is as follows. In Section \ref{sec:KT}, we recall the definition and some properties of bivariate Tratnik polynomials and their relation to Krawtchouk polynomials.
In particular, we provide a new proof of their bispectrality. The $\lambda$-Griffiths polynomials are defined in Section \ref{sec:lG} and their recurrence and difference relations are explicitly given.
Then, Section \ref{sec:bi} provides the biorthogonality relation of these polynomials and Section \ref{sec:as} contains an algebraic interpretation thereof. 
We conclude with some outlooks in Section \ref{sec:out} and collect different properties of the Krawtchouk 
polynomials in Appendix \ref{app:Kraw} .

\section{Properties of Krawtchouk and Tratnik polynomials \label{sec:KT} }
In this section, we recall the construction of bivariate Tratnik polynomials as product of Krawtchouk polynomials, as well as some of their properties. 
We provide an elementary proof of the bispectrality of bivariate Tratnik  polynomials.
\subsection{Krawtchouk polynomials}
We first introduce the Krawtchouk polynomials\footnote{We loosely call $k_i(x;p,N)$ a polynomial while it is a polynomial w.r.t.\ $x$ only up to a normalization.}:
\begin{align}
  k_i(x;p,N)=\left( \frac{p}{1-p}\right)^x \binom{N}{x} \  {}_2F_1 \Biggl({{-i, \;-x}\atop
{-N}}\;\Bigg\vert \;\frac{1}{p}\Biggr)\,,
\,
\end{align}
where $p$ is a free  parameter, while $i$ and $N$ are non-negative integers obeying $0\leq i\leq N$.
One recovers up to a normalization the polynomials defined in \cite{Kra} (see also for example the textbook \cite{Koek}) which are polynomials 
of degree $i$ with respect to the variable $x$ and depending on a parameter $p$. 
The Krawtchouk polynomials are orthogonal and it reads as follows with our choice of normalization:
\begin{align}
 \sum_{x=0}^N w(x;p,N) k_{i}(x;p,N)k_{\ell}(x;p,N)  =   \frac{w(i;p,N)}{(1-p)^N}\ \delta_{i,\ell} \,,\label{eq:ortho}
\end{align}
where
\begin{align}
  &w(x;p,N)=\left(\frac{1-p}{p}\right)^x x!(N-x)!\,.
\end{align}
They satisfy the duality property:
\begin{align} w(x;p,N)\,
 k_i(x;p,N)=
w(i;p,N)\, k_x(i;p,N)\,. \label{eq:duality}
\,
\end{align}
A list of other properties, used in the following, is gathered in Appendix \ref{app:Kraw}.

\subsection{Tratnik polynomials of Krawtchouk type}

We consider here the bivariate Tratnik  polynomials constructed from Krawtchouk polynomials \cite{Trat}. They form a family of bispectral orthogonal polynomials. 
They are defined as follows, for $i,j\geq 0$ and $i+j\leq N$,
\begin{align}\label{eq:tratnik}
 T_{i,j}(x,y;p_1,p_2,N)=k_i(x;p_1,N-j)\,k_j(y;p_2,N-x)\,.
\end{align}
As previously, $T_{i,j}(x,y;p_1,p_2,N)$ is loosely referred as a polynomial although it is a polynomial only up to a normalization.
In this section, we simply write $T_{i,j}(x,y;p_1,p_2,N)=T_{i,j}(x,y)$ to lighten the notation.
We shall prove, by direct computation (\textit{i.e.} using only properties of the Krawtchouk polynomials), 
that the bivariate Tratnik polynomials \eqref{eq:tratnik} satisfy recurrence and difference relations. We recover the results established for example in \cite{GI}.

\begin{prop}\label{pro:recuT}
 The bivariate Tratnik  polynomials $T_{i,j}(x,y)$ satisfy the following recurrence relations:
 \begin{align}
 -x\,T_{i,j}(x,y)&=p_1(N-i-j) T_{i+1,j}(x,y)+i(1-p_1)T_{i-1,j}(x,y)\label{eq:r1}\\
 &-\Big[p_1(N-i-j)+ i(1-p_1) \Big]T_{i,j}(x,y)\,,\nonumber\\
 -y\,T_{i,j}(x,y)&=
  -p_1p_2(N-i-j)T_{i+1,j}(x,y)-i(1-p_1)p_2 T_{i-1,j}(x,y)\nonumber\\
&+p_2(N-i-j) T_{i,j+1}(x,y)+j(1-p_1)(1-p_2)T_{i,j-1}(x,y)\nonumber\\
  &+jp_1(1-p_2)T_{i+1,j-1}(x,y)+i p_2 T_{i-1,j+1}(x,y)\label{eq:r2}\\
 &-\Big[(1-p_1)p_2(N-i-j)+j(1-p_2)+ip_1p_2 \Big]T_{i,j}(x,y)\,.\nonumber
\end{align}
\end{prop}
\proof To prove \eqref{eq:r1}, we use the recurrence relation \eqref{eq:recu} to replace $-x\, k_i(x;p_1,N-j)$ and we recover the L.H.S. of \eqref{eq:r1}. 
To prove relation \eqref{eq:r2}, we start in the same way and use the recurrence relation \eqref{eq:recu} to replace $-y\,k_j(y;p_2,N-x)$ in order to get 
\begin{equation}
\begin{split}
 -y\,T_{i,j}(x,y)=k_i(x;p_1,N-j)&\Big\{ p_2(N-x-j)k_{j+1}(y;p_2,N-x)\\
 &-\big[ p_2(N-x-j)+j(1-p_2) \big] k_{j}(y;p_2,N-x)\\
 & +j(1-p_2)k_{j-1}(y;p_2,N-x)\Big\}\,.
\end{split}
\end{equation}
We then use the contiguity recurrence relation \eqref{eq:contrecu1} to transform $(N-x-j)k_i(x;p_1,N-j)$ in the first line,
the recurrence relation \eqref{eq:recu} to replace $-xk_i(x;p_1,N-j)$ in the second line and the contiguity recurrence relation \eqref{eq:contrecu2}
to transform $k_i(x;p_1,N-j)$ in the third line. This finishes the proof of relation \eqref{eq:r2}.
\endproof

\begin{prop}\label{pro:diffT}
 The bivariate Tratnik  polynomials $T_{i,j}(x,y)$ satisfy the following difference relations:
 \begin{align}
-(j+1)\,T_{i,j}(x,y)&=(1-p_2)(y+1) T_{i,j}(x,y+1)+p_2(N-x-y+1)T_{i,j}(x,y-1)\label{eq:d1}\\
 &-\Big[(1-p_2)(y+1) +p_2(N-x-y+1) \Big]T_{i,j}(x,y)\,,\nonumber\\
-(i+1)\,T_{i,j}(x,y)&=(1-p_1)(x+1) T_{i,j}(x+1,y)+p_1(1-p_2)(N-x-y+1)T_{i,j}(x-1,y) \label{eq:d2}\\
 &-p_1(1-p_2)(y+1) T_{i,j}(x,y+1)-p_1 p_2(N-x-y+1)T_{i,j}(x,y-1)\nonumber\\
 &+\frac{(1-p_1)p_2}{1-p_2}(x+1)T_{i,j}(x+1,y-1) +p_1(1-p_2)(y+1)T_{i,j}(x-1,y+1) \nonumber\\
 &-\Big[(1-p_1)(x+1)+p_1p_2(y+1)  +p_1(1-p_2)(N-x-y+1) \Big]T_{i,j}(x,y)\,.\nonumber
\end{align}
\end{prop}
\proof The proof follows the same lines as the proof of Proposition \ref{pro:recuT} by using the difference relation \eqref{eq:diff} 
and the contiguity difference relations \eqref{eq:contdiff1} and \eqref{eq:contdiff2} instead of 
the recurrence and contiguity recurrence relations. \endproof
Similarly, using only the orthogonality property of Krawtchouk polynomials, one can show that
bivariate Tratnik  polynomials \eqref{eq:tratnik} are orthogonal for the trinomial distribution.

\section{Bispectrality of the $\lambda$-Griffiths polynomials \label{sec:lG}}

We now introduce a generalization of the bivariate Griffiths  polynomials \cite{Griff}, that we call
 $\lambda$-Griffiths polynomials:
\begin{equation}
\begin{split}
G^\lambda_{i,j}(x,y;p_1,p_2,p_3;N)&=\sum_{a=0}^{N-j} \lambda^a\, k_i(a;p_1,N-j)\,k_j(y;p_2,N-a)\,k_a(x;p_3,N-y)\,,
\end{split}
\end{equation}
where $p_1,p_2,p_3$ and $\lambda$ are free parameters. At the end of the paper, we show that for a particular value of $\lambda$, one recovers the usual bivariate Griffiths  polynomials. 
When there is no ambiguity, we use the shortened notation $G^\lambda_{i,j}(x,y)=G^\lambda_{i,j}(x,y;p_1,p_2,p_3;N)$.
The  $\lambda$-Griffiths polynomials can be rewritten in terms of the bivariate Tratnik  polynomials as
\begin{align}
G^\lambda_{i,j}(x,y) &=\sum_{a=0}^{N-j} \lambda^a\, T_{i,j}(a,y;p_1,p_2,N)\,k_a(x;p_3,N-y)=\sum_{a=0}^{N-j} \lambda^a\, k_i(a;p_1,N-j)\,T_{ja}(y,x;p_2,p_3,N)\,.\label{eq:TG}
\end{align}
Let us mention that the $\lambda$-Griffiths polynomials also satisfy a duality relation.
It reads 
\begin{equation}\label{eq:dualiltyG}
\begin{split}
 \widetilde G_{i,j}^\lambda(x,y;p_1,p_2,p_3;N) = \widetilde G_{x,y}^\lambda(i,j;p_3,p_2,p_1;N)\,,
\end{split}
\end{equation}
where
\begin{equation}\label{eq:dualiltyGa}
 \widetilde G_{i,j}^\lambda(x,y;p_1,p_2,p_3;N) =\frac{1}{i!j!(N-i-j)!}  \left(\frac{p_1}{1-p_1}\right)^i \left(\frac{p_2}{1-p_2}\right)^{j}\,
 G_{i,j}^{\lambda \frac{1-p_1}{p_1}}(x,y;p_1,p_2,p_3;N) \,.
\end{equation}
It can be deduced from the duality \eqref{eq:duality} of the Krawtchouk polynomials.


The $\lambda$-Griffiths polynomials are bispectral as shown in the following propositions.
\begin{prop}\label{pro:recuG}
The $\lambda$-Griffiths polynomials $G^\lambda_{i,j}(x,y)$ satisfy the following recurrence relations:
\begin{align}
 -x\,G^\lambda_{i,j}(x,y)&= 
\frac{1-p_1}{\lambda} \left(\frac{\lambda p_1( 1-p_3)}{1-p_1}-p_3\right)\left(\frac{\lambda p_1(1-p_2)}{1-p_1}+1\right)(N-i-j)G^\lambda_{i+1,j}(x,y)\label{eq:recuG1}\\
 -&\frac{1-p_1}{\lambda} \left(\lambda(1-p_3)+p_3\right)(\lambda(1- p_2)-1)  i\,G^\lambda_{i-1,j}(x,y)\nonumber\\ 
 +&p_2\left(\frac{\lambda p_1(1-p_3)}{1-p_1}-p_3 \right)(N-i-j)\,G^\lambda_{i,j+1}(x,y) 
-\frac{(1-p_1)p_3}{\lambda}\left(\lambda(1-p_2)-1\right) j\,G^\lambda_{i,j-1}(x,y) \nonumber\\
 -& \frac{p_3}{\lambda}\left( \lambda p_1(1-p_2)+1-p_1\right) j\, G^\lambda_{i+1,j-1}(x,y)
-p_2(\lambda(1-p_3)+p_3) i\,  G^\lambda_{i-1,j+1}(x,y)\nonumber\\ 
 -&\Big[ \frac{(\lambda (1-p_3)+p_3)(\lambda p_1(1-p_2)+1-p_1)}{\lambda} i+p_2p_3 j\nonumber\\
 &\qquad -\frac{(\lambda (1-p_2)-1)(\lambda p_1( 1-p_3)-p_3(1-p_1))}{\lambda}(N-i-j)\Big]\, G^\lambda_{i,j}(x,y)\,, \nonumber\\
 -y\,G^\lambda_{i,j}(x,y)&=-p_1p_2 (N-i-j)G^\lambda_{i+1,j}(x,y)  -p_2(1-p_1)\,  i\,G^\lambda_{i-1,j}(x,y) \label{eq:recuG2}\\
 &   +p_2(N-i-j)G^\lambda_{i,j+1}(x,y)+(1-p_1)(1-p_2) j\,G^\lambda_{i,j-1}(x,y) \nonumber\\
&+  p_1(1-p_2)\, j\, G^\lambda_{i+1,j-1}(x,y)+p_2 i\,  G^\lambda_{i-1,j+1}(x,y)\nonumber\\
 & -\Big[ p_1p_2i+(1-p_2) j+p_2(1-p_1)(N-i-j)\Big]\, G^\lambda_{i,j}(x,y)\nonumber\,.
\end{align}
\end{prop}
\proof To prove \eqref{eq:recuG1}, let us express the $\lambda$-Griffiths polynomial in terms of the Tratnik ones using the second expression in \eqref{eq:TG}.
Then, recurrence relation \eqref{eq:r2} for $-x\, T_{ja}(y,x;p_2,p_3,N)$ leads to
\begin{align}
 -x\,G^\lambda_{i,j}(x,y)&= \sum_{a}\lambda^a k_i(a;p_1,N-j) \left\{ \sum_{(\epsilon,\epsilon')\in \cW}C^{(\epsilon,\epsilon')}_{j,a} T_{j+\epsilon, a+\epsilon'}(y,x;p_2,p_3,N)  \right\}\\
 &= \sum_{(\epsilon,\epsilon')\in \cW} \sum_{a}\lambda^a\ C^{(\epsilon,\epsilon')}_{j,a} k_i(a;p_1,N-j)\  T_{j+\epsilon, a+\epsilon'}(y,x;p_2,p_3,N)  \,,
 \end{align}
where $\cW=\{(0,0),(1,0),(-1,0),(0,1),(0,-1),(1,-1),(-1,1)\}$ and $C^{(\epsilon,\epsilon')}_{j,a}$ are the coefficients appearing in the recurrence relation \eqref{eq:r2}.
Now, we use different relations given in Appendix \ref{app:Kraw} depending on $(\epsilon,\epsilon')$ to transform the coefficients $C^{(\epsilon,\epsilon')}_{j,a} k_i(a;p_1,N-j)$ in order to recognize $\lambda$-Griffiths polynomials.
More precisely, we use
\begin{itemize}
 \item[\textit{(i)}] Recurrence relation \eqref{eq:recu} for $(\epsilon,\epsilon')=(0,0)$;
 \item[\textit{(ii)}] Contiguity recurrence relation \eqref{eq:contrecu1} for $(\epsilon,\epsilon')=(1,0)$;
 \item[\textit{(iii)}] Contiguity recurrence relation \eqref{eq:contrecu2} for $(\epsilon,\epsilon')=(-1,0)$;
 \item[\textit{(iv)}] Shifted recurrence relation \eqref{eq:sr2} for $(\epsilon,\epsilon')=(0,1)$;
 \item[\textit{(v)}] Shifted recurrence relation \eqref{eq:sr1} for $(\epsilon,\epsilon')=(0,-1)$;
 \item[\textit{(vi)}] Dual backward relation \eqref{eq:scr2} for $(\epsilon,\epsilon')=(1,-1)$;
 \item[\textit{(vii)}] Dual forward relation \eqref{eq:scr1} for $(\epsilon,\epsilon')=(-1,1)$.
\end{itemize}
Let us treat the case $(iv)$ in detail. In this case, one has to compute 
\begin{align}
& \sum_{a=0}^{N-j} \lambda^a\ C^{(0,1)}_{j,a} k_i(a;p_1,N-j)\  T_{j, a+1}(y,x;p_2,p_3,N)\nonumber\\
&=p_3\sum_{a=0}^{N-j} \lambda^a\ (N-j-a) k_i(a;p_1,N-j)\  T_{j, a+1}(y,x;p_2,p_3,N)\nonumber\\
 &=p_3\sum_{a=1}^{N-j} \lambda^{a-1}\ (N-j-a+1) k_i(a-1;p_1,N-j)\  T_{j, a}(y,x;p_2,p_3,N)\nonumber\\
 &=p_3\sum_{a=1}^{N-j} \lambda^{a-1}\ \Big\{ (i+j-N) k_{i+1}(a;p_1,N-j)+i k_{i-1}(a;p_1,N-j) \nonumber\\
 &\hspace{3cm}+(N-j-2i) k_{i}(a;p_1,N-j)\Big\}  T_{j, a}(y,x;p_2,p_3,N)\,.\nonumber
\end{align}
We have used the shifted recurrence relation \eqref{eq:sr2} in the last step.
Since the factor in brackets in the last relation vanishes for $a=0$ (we used $k_j(0;p,N)=1$), the sum over $a$ can start from $a=0$.
 Therefore we can recognize $\lambda$-Griffiths polynomials. The other cases are treated similarly, hence \eqref{eq:recuG1} is proven.
 
 For relation \eqref{eq:recuG2}, we express the $\lambda$-Griffiths polynomial in terms of the Tratnik one using the first expression in \eqref{eq:TG}.
Then, the recurrence relation \eqref{eq:r2} for $-yT_{i,j}(a,y;p_1,p_2,N)$ allows to prove \eqref{eq:recuG2}. \endproof

\begin{prop}\label{pro:diffG}
 The $\lambda$-Griffiths polynomials $G^\lambda_{i,j}(x,y)$ satisfy the following difference relations:
\begin{align}
 -i\, G^\lambda_{i,j}(x,y)&= (x+1)\ \frac{1-\lambda}{\lambda}(1-p_3)(\lambda p_1( p_2-1)+p_1-1) G^\lambda_{i,j}(x+1,y)\label{eq:diffG1} \\
 -&(N-x-y+1)\ \frac{(\lambda (p_3-1)-p_3)}{\lambda(p_3-1)}(\lambda p_1 (1-p_2) (1-p_3)-p_3(1-p_1))G^\lambda_{i,j}(x-1,y) \nonumber\\
 -&(1+y)\ p_1(1-p_2)(1-\lambda)G^\lambda_{i,j}(x,y+1)\nonumber\\
 -&(N-x-y+1)\ \frac{p_2}{\lambda (1-p_2)}(\lambda p_1 (1-p_2)(1-p_3)-p_3(1-p_1) ) G^\lambda_{i,j}(x,y-1)\nonumber\\
 -&(x+1)\ \frac{p_2(1-p_3)}{\lambda (1-p_2)}(\lambda p_1 (1-p_2)+1-p_1) G^\lambda_{i,j}(x+1,y-1)\nonumber\\
 -&(y+1) \ \frac{p_1(1-p_2)}{1-p_3}(\lambda (1-p_3)+p_3) G^\lambda_{i,j}(x-1,y+1)\nonumber\\
 -&\Big[x\ \frac{1}{\lambda}(\lambda (1-p_3)+p_3)(\lambda p_1(1-p_2)+1-p_1)  +y\ p_1 p_2\nonumber\\
 &\qquad +(N-x-y)\frac{1-\lambda}{\lambda}(\lambda p_1 (1-p_2)(1-p_3)-p_3(1-p_1) ) \Big] G^\lambda_{i,j}(x,y)\,,\nonumber
\end{align}
 \begin{align}
 -j \,G^\lambda_{i,j}(x,y)&=-(x+1)\ p_2(1-p_3) G^\lambda_{i,j}(x+1,y)  -(N-x-y+1) \ p_2p_3G^\lambda_{i,j}(x-1,y) \label{eq:diffG2}\\
 &   +(y+1)\ (1-p_2)G^\lambda_{i,j}(x,y+1)+(N-x-y+1)\ p_2(1-p_3)  G^\lambda_{i,j}(x,y-1) \nonumber\\
&+(x+1) \  p_2(1-p_3) G^\lambda_{i,j}(x+1,y-1)+(y+1)\ (1-p_2)\frac{p_3}{1-p_3}   G^\lambda_{i,j}(x-1,y+1)\nonumber\\
 & -\Big[ x\ p_3p_2+y\ (1-p_2) +(N-x-y)\ p_2(1-p_3)\Big] G^\lambda_{i,j}(x,y)\nonumber\,.
\end{align}
\end{prop}
\proof The proof can be done following the same lines as previously for Proposition \ref{pro:recuG}. However, a simpler approach is to use 
the duality relation \eqref{eq:dualiltyG}, replace each $\lambda$-Griffiths polynomial in the result of Proposition \ref{pro:recuG} and  
rename $(i,j,x,y)\to (x,y,i,j)$ to prove the result of the proposition. \endproof

\section{Biorthogonality of the $\lambda$-Griffiths polynomials \label{sec:bi}}

$\lambda$-Griffiths polynomials satisfy a more general relation, called biorthogonality relation, given in the next proposition.
\begin{prop}
 The $\lambda$-Griffiths polynomials $G^\lambda_{i,j}(x,y)$ satisfy a biorthogonality relation given explicitly by
\begin{align}
\sum_{x,y} \Omega_{x,y}(p_2,p_3)\  G^\lambda_{i,j}(x,y)\ G^{\mu/ \lambda}_{i',j'}(x,y)  =\frac{ \Omega_{i,j}(p_2,p_1)}{(1-p_1)^N(1-p_2)^N(1-p_3)^N}\ \delta_{i,i'} \delta_{j,j'}\,, \label{eq:bio}
\end{align}
where 
\begin{align}
&\mu=\frac{(1-p_1)p_3}{p_1(1-p_2)(1-p_3)}\, , \\
 &\Omega_{x,y}(p_1,p_2)= x!y!(N-x-y)!  \ \left( \frac{1-p_1}{p_1} \right)^y \left( \frac{1-p_2}{p_2} \right)^x (1-p_2)^y  \, .
\end{align}
\end{prop}
\proof It is a direct check to prove \eqref{eq:bio} using the orthogonality relation \eqref{eq:ortho} of the Krawtchouk polynomials. \endproof

Let us remark that the weight in the biorthogonality relation does not depend on $\lambda$.
For $\lambda^2= \mu= \frac{(1-p_1)p_3}{p_1(1-p_2)(1-p_3)}$, the $\lambda$-Griffiths polynomials become orthogonal. 
For the case $\lambda= +\sqrt{\frac{(1-p_1)p_3}{p_1(1-p_2)(1-p_3)}}$, we recover the bivariate Griffiths  polynomials constructed and studied in \cite{Griff,IT,GVZ}.
To the best of our knowledge, the case $\lambda= - \sqrt{\frac{(1-p_1)p_3}{p_1(1-p_2)(1-p_3)}}$ is new.

\section{Algebraic interpretation of $\lambda$-Griffiths polynomials \label{sec:as}} 
As mentioned previously,  the usual bivariate Griffiths  polynomials have been interpreted in \cite{GVZ} as the entries of a representation of $SO(3)$.
More precisely, let us consider three oscillator algebras generated by $a_i$ and $a_i^\dagger$ for $i=1,2,3$ and satisfying, for $i,j=1,2,3$,
\begin{equation}
 [a_i,a_j]=0\,,\qquad [a^\dagger_i,a^\dagger_j]=0\,,\qquad [a_i,a^\dagger_j]=\delta_{ij}\,.
\end{equation}
For a given $N$, they act on $|n_1,n_2,N-n_1-n_2\rangle$ the standard basis (with $n_1+n_2\leq N$).
Then, taking an element $R=\exp(B)\in SO(3)$ (with $B$ an antisymmetric matrix $B^t=-B$), we can form the following unitary matrix
\begin{equation}
 U(R)=\exp \left( \sum_{i,k=1}^3 B_{ik} a_i^\dagger a_k\right)\,.
\end{equation}
Decomposing this rotation with three Euler angles, one gets
$R=R_{xz}(\phi)R_{yz}(\theta)R_{xz}(\psi)$. The entries of $U(R)=U(R_{xz}(\phi))U(R_{yz}(\theta))U(R_{xz}(\psi))$ in the basis $|n_1,n_2,N-n_1-n_2\rangle$ are given in terms of the usual bivariate Griffiths polynomials
(the parameters $p_1$, $p_2$ and $p_3$ of the polynomials being associated to the three Euler angles).

To obtain an analogous interpretation for the $\lambda$-Griffiths polynomials, let us introduce 
\begin{equation}
U_\varphi=\exp(i\varphi a_1^\dagger a_1)\quad\mbox{where}\quad  \lambda=e^{i\varphi}\sqrt{\mu}.
\end{equation}
Remark that although  $U_\varphi$ is not associated to an element of $SO(3)$, it is still an element of $U(3)$ if $\varphi$ is real.
If $\varphi$ is not real (in particular if we want that $\lambda$ be real), the transformation $U_\varphi$ is not  unitary anymore.
By using the same types of computations as the ones done in \cite{GVZ}, we can show that the entries of
\begin{equation}
 U(R_{xz}(\phi))\, U(R_{yz}(\theta))\, U_\varphi\, U(R_{xz}(\psi))\,,
\end{equation}
in the basis $|n_1,n_2,N-n_1-n_2\rangle$ are given by the $\lambda$-Griffiths polynomials.
When $\varphi=0$, we recover the original construction.

\section{Outlooks \label{sec:out}}

In this paper, we provide a new proof of the bispectrality of the bivariate Griffiths polynomials using only properties of
univariate polynomials. This proof is elementary and allows us to obtain a new generalization of these well-known polynomials that still possesses
bispectral properties but is biorthogonal. We believe that similar proofs will be very useful to study other multivariate polynomials. Indeed, Griffiths-like extensions not
based on Krawtchouk polynomials but on other families of the Askey scheme are scarcely studied.  This paper is a first step towards a general approach to such Griffiths-like polynomials. 
We intend to continue this framework in the near future. Similarly, increasing the number of variables (going from bivariate to multivariate polynomials) is certainly a line of research of great interest. 
In that perspective, a point to clarify would be 
whether the list of properties of univariate polynomials detailed in Appendix \ref{app:Kraw} is necessary and/or sufficient. 

Let us notice that to any bispectral problem, we can associate an underlying algebra. For the univariate case, this leads to the Askey--Wilson algebra and its degenerations \cite{Zhe91,avatar}. 
For the bivariate case, we need a rank 2 algebra. This type of algebra has been used recently to study bivariate Griffiths polynomials of Racah type \cite{icosi}. However, despite the existence of more general rank $n$ algebras,
such as the rank $n$ Racah algebra \cite{DBGV,Post,CGPV} or the rank $n$ Askey--Wilson algebra \cite{DBDCVV, CFPR}, 
their relation to orthogonal multivariate polynomials needs to be explored in more details. 
Let us also mention that algebras associated to the bispectrality of some biorthogonal functions are being 
investigated \cite{VZ_Hahn,VZ_Ask} and prove to be very useful for characterizing the properties of these functions (as well as their orthogonal polynomial counterparts). 

Finally, as explained previously, an algebraic interpretation of  Griffiths polynomials of Krawtchouk type in terms of oscillator algebras has been been given in \cite{GVZ}. 
A similar interpretation involving $q$-oscillator algebras has been obtained for Tratnik polynomials of $q$-Krawtchouk type \cite{GPV,BKV}. It would be interesting to generalize these approaches to more general cases.

\subsection*{Acknowledgements: }
N.~Cramp\'e, J. Gaboriaud and M. Zaimi thank LAPTh for its hospitality. N.~Cramp\'e and M.~Zaimi are partially supported by the international research project AAPT of the CNRS.
L.~Frappat and E.~Ragoucy are partially supported by Universit\'e Savoie Mont Blanc and Conseil Savoie Mont Blanc grant APOINT. 
J.~Gaboriaud is supported by JSPS KAKENHI Grant Numbers 22F21320 and 22KF0189.
The research of L. Vinet is supported by a Discovery Grant from the Natural Sciences and Engineering Research Council (NSERC) of Canada. 
M.~Zaimi holds an Alexander-Graham-Bell scholarship from the NSERC.

\begin{appendix}

\section{Properties of the Krawtchouk polynomial\label{app:Kraw}}

\paragraph{Difference relation}

\begin{align}
-(n+1) k_n(x;p,N)&=(1-p)(x+1) k_n(x+1;p,N)+p(N-x+1)k_n(x-1;p,N)\label{eq:diff}\\
 &-\Big[(1-p)(x+1)+p(N-x+1)\Big]k_n(x;p,N)\,.\nonumber
\end{align}
Up to the normalization, this relation is given in \cite{Koek}.

\paragraph{Recurrence relation}

\begin{align} 
-x\,k_n(x;p,N)&=p(N-n)k_{n+1}(x;p,N) + n(1-p)k_{n-1}(x;p,N) \label{eq:recu}\\
&-\Big[p(N-n) +n(1-p)\Big]k_{n}(x;p,N)\,.\nonumber
\end{align}
Up to the normalization, this relation is given in \cite{Koek}.

\paragraph{Contiguity difference relation (I)}

\begin{align} 
& k_n(x;p,N)= k_{n}(x;p,N-1) +\frac{p}{1-p}k_{n}(x-1;p,N-1)\,. \label{eq:contdiff1}
\end{align}
This type of relation has been proven in the case of the Racah polynomials in \cite{icosi}. We reproduce the proof here for the Krawtchouk case.
This relation is proven recursively on $x$. For $x=1$, it is proven by direct computation.
Let us suppose that \eqref{eq:contdiff1} is valid up to $x$. Then, let us express $k_n(x+1;p,N)$
using the difference relation \eqref{eq:diff}
\begin{align} 
& k_n(x+1;p,N)=\left[-\frac{n+1}{(1-p)(x+1)}+1+\frac{p(N-x+1)}{(1-p)(x+1)}\right] k_{n}(x;p,N)-\frac{p(N-x+1)}{(1-p)(x+1)}k_{n}(x-1;p,N)\,. \nonumber
\end{align}
Using the recurrence hypothesis, one can express $k_{n}(x;p,N)$ and $k_{n}(x-1;p,N)$ using relation \eqref{eq:contdiff1} to get 
\begin{align} 
 k_n(x+1;p,N)&=\left[-\frac{n+1}{(1-p)(x+1)}+1+\frac{p(N-x+1)}{(1-p)(x+1)}\right]  \left( k_{n}(x;p,N-1) +\frac{p}{1-p}k_{n}(x-1;p,N-1)\right)\nonumber\\
&-\frac{p(N-x+1)}{(1-p)(x+1)}\left( k_{n}(x-1;p,N-1) +\frac{p}{1-p}k_{n}(x-2;p,N-1)\right)\,. \nonumber
\end{align}
We use again difference relation \eqref{eq:diff} to express $-(n+1)k_{n}(x;p,N-1)$ and $-(n+1)k_{n}(x-1;p,N-1)$.
By regrouping the terms, we prove that \eqref{eq:contdiff1} is valid for $x+1$ which finishes the proof of \eqref{eq:contdiff1}.

\paragraph{Contiguity difference relation (II)} 
\begin{align} 
& \frac{N+1-n}{1-p}k_n(x;p,N)= (N+1-x) k_{n}(x;p,N+1) +(x+1)k_{n}(x+1;p,N+1)\,. \label{eq:contdiff2}
\end{align}
This relation is proven following the same lines as the proof of the previous relation.

\paragraph{Contiguity recurrence relation (I)}

\begin{align}
& (N-x)k_n(x;p,N)=(N-n)k_n(x;p,N-1)+n k_{n-1}(x;p,N-1)\,. \label{eq:contrecu1}
\end{align}
To prove this relation, we use the duality \eqref{eq:duality} on the contiguity difference relation \eqref{eq:contdiff1} then we rename $x$ and $n$: $(x,n)\to(n,x)$.

\paragraph{Contiguity recurrence relation (II)}
\begin{align}
&  k_n(x;p,N)=pk_{n+1}(x;p,N+1)+(1-p) k_{n}(x;p,N+1)\,.\label{eq:contrecu2}
\end{align}
To prove this relation, we use the duality \eqref{eq:duality}  on the contiguity difference relation \eqref{eq:contdiff2} then we rename $x$ and $n$: $(x,n)\to(n,x)$.

\paragraph{Forward relation}

\begin{align} 
& \frac{n+1}{1-p}k_n(x;p,N)=  \frac{p}{1-p}(N+1-x)k_{n+1}(x;p,N+1) -(x+1) k_{n+1}(x+1;p,N+1)\,. \label{eq:for}
\end{align}
Up to the normalization, this relation is given in \cite{Koek}.

\paragraph{Backward relation}

\begin{align} 
& k_n(x;p,N)=  k_{n-1}(x;p,N-1) -k_{n-1}(x-1;p,N-1)\,.\label{eq:back}
\end{align}
Up to the normalization, this relation is given in \cite{Koek}.

\paragraph{Dual forward relation}

\begin{align}
 &  \frac{1}{1-p}k_n(x;p,N)= k_{n}(x+1;p,N+1)-k_{n+1}(x+1;p,N+1)\,.\label{eq:scr1}
\end{align}
To prove this relation, we use the duality \eqref{eq:duality} on the forward relation then we rename $x$ and $n$: $(x,n)\to(n,x)$.

\paragraph{Dual backward relation}

\begin{align}
& x\,k_n(x;p,N)=\frac{p(N-n)}{1-p}k_{n}(x-1;p,N-1)-n k_{n-1}(x-1;p,N-1)\,.\label{eq:scr2}
\end{align}
To prove this relation, we use the duality \eqref{eq:duality} on the backward relation then we rename $x$ and $n$: $(x,n)\to(n,x)$.

\paragraph{Shifted difference relations (I)}

\begin{align} 
 \frac{N+1-n}{1-p}k_n(x;p,N)&= (x+1) k_{n-1}(x+1;p,N)-(N+1-x)k_{n-1}(x-1;p,N)\nonumber\\
 &+(N-2x)k_{n-1}(x;p,N)\,.
\end{align}
To prove this relation, we replace the two Krawtchouk polynomials on the R.H.S. of the backward relation \eqref{eq:back} by using the contiguity relation \eqref{eq:contdiff2}.

\paragraph{Shifted difference relations (II)}

\begin{align} 
 (n+1)k_n(x;p,N)&= -(1-p)(x+1) k_{n+1}(x+1;p,N) +\frac{p^2}{1-p}(N+1-x)k_{n+1}(x-1;p,N)\nonumber\\
 &-p(2x-N)k_{n+1}(x;p,N)\,.
\end{align}
To prove this relation, we replace the two Krawtchouk polynomials on the R.H.S. of the forward relation \eqref{eq:for} by using the contiguity relation \eqref{eq:contdiff1}.

\paragraph{Shifted recurrence relations (I)}

\begin{align}
 x\,k_n(x;p,N)&= \frac{p^2(N-n)}{1-p} k_{n+1}(x-1;p,N)-n(1-p)k_{n-1}(x-1;p,N)\nonumber\\
 & +p(N-2n)k_{n}(x-1;p,N)\,. \label{eq:sr1}
\end{align}
To prove this relation, we replace the two Krawtchouk polynomials on the R.H.S. of the dual backward relation \eqref{eq:scr2} by using the contiguity relation \eqref{eq:contrecu2}.

\paragraph{Shifted recurrence relations (II)}

\begin{align}
 \frac{N-x}{1-p}k_n(x;p,N)&= (n-N) k_{n+1}(x+1;p,N) +nk_{n-1}(x+1;p,N)\nonumber\\
 &+(N-2n)k_{n}(x+1;p,N)\,.\label{eq:sr2}
\end{align}
To prove this relation, we replace the two Krawtchouk polynomials on the R.H.S. of the dual forward relation \eqref{eq:scr1} by using the contiguity relation \eqref{eq:contrecu1}.

\end{appendix}

\end{document}